\documentclass[reprint,amsmath,amssymb,aps]{revtex4-1}
\usepackage{graphicx}
\usepackage{epsfig}
\usepackage{float}
\usepackage{hyperref}
\usepackage{amsmath}
\usepackage{amssymb}
\usepackage{graphics}
\usepackage{setspace}
\newcommand{\mathsym}[1]{{}}
\newcommand{\unicode}[1]{{}}
\newcommand{\abs}[1]{\left| #1 \right|}

\begin{document}
\preprint{AIP/PRA}
\title{An exact analytical scheme using a new potential to solve one-dimensional quantum systems}
\author{Rajendran Saravanan}
\email{saravanan.quark@gmail.com}
\homepage{\\https://saravananrajendran.weebly.com/}
\author{Deepak Kumar}
\author{Aniruddha Chakraborty}
\affiliation{School of Basic Sciences, Indian Institute of Technology Mandi, Mandi-175075, India.}
\date{\today}
\begin{abstract}
\noindent We propose an exact method for solving a one-dimensional Schr\"odinger equation. An arbitrary potential is represented by the collection of short-width potentials. For building the collection scheme, a new solvable potential is introduced. It is based on the simple expansion of the wavefunction of the introduced potential. The illustration of the scheme is done by reproducing the results of the rectangular potential. The scheme has computational advantages and the transmission properties, eigenenergies can be calculated efficiently. The presented scheme is compared with the other similar schemes in terms of computational complexity, analytical solubility, etc.. A \textit{Mathematica} code is provided in the supplementary file that solves the Schr\"odinger equation with arbitrary potential function $V(x)$ and effective mass $m(x)$.
\end{abstract}
\maketitle
\section{Introduction}
\noindent Finding exact analytical solutions for time-independent Schr\"odinger equation can improve understanding over several atomic, molecular and nano-level systems \cite{balantekin1998quantum,gamow1928quantentheorie,forbes2011transmission,datta1990quantum,luscombe1990models,harris1989electronic,kronig1931quantum,domany1983solutions,chang1974resonant,cahay1989electron}. Some solvable models have been used in the literature that validates experiments and the theoretical hypotheses \cite{osorio1988bound, gamow1928quantentheorie, kronig1931quantum}. But in reality, the potential profiles have complicated functional form. Therefore, it is of significance to introduce new solvable potentials \cite{gerber1974properties} with new analytical schemes to solve arbitrary one-dimensional systems. The analytical schemes has been given in the literature for the existing analytical models. Such schemes involve 
piece-wise constant \cite{gilmore2004elementary,ando1987calculation}, piece-wise linear \cite{lui1986exact}, collection of Dirac delta functions \cite{chakraborty2009multi}, collection of rectangular potential functions \cite{gilmore2004elementary} to solve an arbitrary potential \cite{gilmore2004elementary}. Some of the above schemes involve complicated wavefunctions and hence the expansion becomes inefficient. In the paper, we propose a new solvable potential with an ultra-short width. An analytical scheme is built using the collection of wavefunctions that utilizes the transfer-matrix technique. The simple form of the wavefunction make the algorithm have computational or analytical advantage. The transmission probability, eigenenergies and the eigenfunctions can be calculated efficiently using the algorithm. 
The paper is organised as follows: section II derives the bound state properties, transmission probability, and Ramsauer probability of the new potential. Section III discusses the properties of the ultra-short potential. Section IV illustrates the solubility in the time-dependent framework by deriving the Laplace transformed wavefunction $\psi(x,s)$ of the model potential. The proposed model can be solved for arbitrary asymptotic potential energy regions and arbitrary initial wavepacket. Section V presents the transfer matrix scheme for collecting the ultrashort potentials. In section VI, the analytical scheme is validated by reproducing the results of rectangular potential function. The paper is concluded by comparing the features of analytical schemes with other schemes. The model potential has advantages of having less computational complexity and because of the solubility in the time-domain \cite{saravanan2019exact}. We present a \textit{Mathematica} code in the supplementary, that can discretize an arbitrary potential and present the quantum properties. We conclude by discussing the scope and improvements over the present work.
\section{Quantum properties of the ultra-short potential}
\subsection{Bound states of the ultra-short potential well}
\noindent The time-independent Schr\"odinger equation for an arbitrary potential well is written as,
\begin{equation}
-\frac{\hbar^2}{2m}\frac{d^2\psi}{d x^2}-V(x)\psi=E\psi,
\label{eqn:tise}
\end{equation}
\noindent where $E$ is the energy of the quantum object, $m$ is the mass, and $V(x)$ is the potential energy function. When the object is assumed to encounter an ultra-short potential given by,
\begin{equation}
V(x)=\left\{ \begin{array}{ll}
V(x) & \mbox{if $\abs{x}<\delta x$}\\
0 & \mbox{otherwise},
\end{array}\right.
\label{eq1}
\end{equation} 
where `$\delta x$' represents ultra-short width. 
\begin{figure}
\includegraphics[scale=0.5]{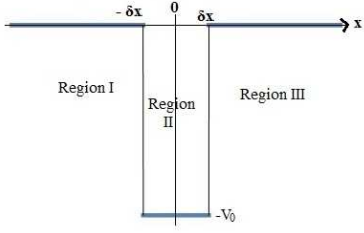}
\caption{Schematic representation of the ultra-short potential well}
\label{fig:ultra}
\end{figure}
\noindent When the potential has an ultra-short width, the variation can be taken to be a zeroth order expansion i.e., $V(x)=V_0$ in the region II (Fig. \ref{fig:ultra}). The corresponding wave function can also be assumed to be curvature-less inside that region only $\psi_{II}(x)=\psi(0)$. Now, this is a crucial step that aids computational advantage by having a simpler form for the wavefunction. 
\noindent The Eq. \eqref{eqn:tise} has the following solution corresponding to the piece-wise energy regions for the case $E<0$,
\begin{equation}
\psi=\begin{cases}
Ae^{\kappa x} & x<-\delta x\\
\psi_c(0) & x\in (-\delta x,\delta x)\\
Be^{-\kappa x} & x>\delta x,~~~~~~~~~\text{where}~\kappa^2=-2mE/\hbar^2.
\end{cases}
\label{eqn:phi}
\end{equation}
\noindent Using the wave function continuity at $x= 0 - \delta x$ and $x = 0 + \delta x$ we get,
\begin{eqnarray}
B e^{-\kappa\delta x}=\psi_c(0), \\ \nonumber
A e^{-\kappa\delta x}=\psi_c(0).
\end{eqnarray}
\noindent We derive the boundary conditions using the standard method of integrating the time independent Schr\"odinger equation from $-\delta x-\epsilon$ to $\delta x+\epsilon$ in the limit $\epsilon\to0$ to get
\begin{equation}
\left(\frac{\partial \psi}{\partial x}\right)_{-\delta x}^{\delta x}=-(E+
V_0)\psi_c(0)\frac{4m\delta x}{\hbar^2}.
\label{eqn:newbc}
\end{equation}
\noindent It is worthwhile to note that the presence of the energy E in the boundary condition make the proposed model different than the rest analytical models. The use of the above boundary condition gives the following equation, which may be used to estimate the bound state energy
\begin{equation}
\frac{\sqrt{-2mE}}{\hbar}=\frac{2m}{\hbar^2}\delta x (E+V_0),\label{eqn:mo}
\end{equation}
\noindent solving which yields,
\begin{equation}\implies E_{+}=\frac{\sqrt{8 \delta x^2 m V_0 \hbar ^2+\hbar ^4}-4 \delta x^2 m
V_0-\hbar ^2}{4 \delta x^2 m}.
\label{eqn:Epm}
\end{equation}
We choose the positive root of Eq. \ref{eqn:newbc} which otherwise would lead to an unphysical bound energy that is lesser than $-V_0$. The normalization constant is found to be the following,
\begin{equation}
B= \frac{e^{\kappa\delta x}}{\sqrt{1/\kappa+2\delta x}}.\label{eqn:normconst}
\end{equation}
\noindent Eqs. \ref{eqn:Epm}, \ref{eqn:phi} suggest that the ultra-short well has single bound state and the eigen-function will be of even parity. The even parity solution suggests that it is the result of a symmetric nature of the ultra-short potential. The bound state wave function of the Dirac delta potential well is given by \cite{griffiths2016introduction},
\begin{equation}
\psi(x)=\sqrt{\kappa}e^{-\kappa\abs{x}},
\end{equation}
\noindent where $\kappa$ is defined as $\kappa^2=-\frac{2mE}{\hbar^2}$ supporting one allowed value known to be,
\begin{equation}
\kappa=\frac{m k_0}{\hbar^2}.
\end{equation}
\noindent The energy of that bound state is given by
\begin{equation}
E = - \frac{m k_0^2}{2 \hbar^2}.
\end{equation}Both the potential well, ultra-short and Dirac delta supports only one bound state, but the energy value and the corresponding eigen-function of the ultra-short is very different from that of the Dirac delta function.
\subsection{Transmission coefficient for ultra-short potential barrier}
\noindent The energy eigenvalue equation in presence of a potential barrier is given as,
\begin{equation}
-\frac{\hbar^2}{2m}\frac{d^2\phi}{dx^2}+V(x)\phi=E\phi,
\end{equation}
\noindent with $V(x)>0$. The barrier is non-zero in an ultra-short region given by $x\in[-\delta x,\delta x]$. Hence, the potential can be written as zero-th order expansion around a point {\it i.e.,} $V(x)=V(0)=V_0$ and so the wavefunction in the small region as $\phi_{II}=\phi_c(0)$. Solving similarly the piece-wise regions for $E>0$ yields the following wavefunction,
\begin{equation}
\phi=\begin{cases}
Ae^{ik_1x}+Be^{-ik_1x}&x<-\delta x\\
\phi_c(0)&-\delta x<x<\delta x\\
Ce^{ik_1x}&x>\delta x~~~\text{where}~~k_1^2=\frac{2mE}{\hbar^2}.
\end{cases}
\end{equation}
The wavefunction obeys the following continuity and derivative requirements at the boundaries $x=\pm\delta x$
\begin{eqnarray} \phi(x)|_{x=-\delta x}=\phi_c(0), \\ \nonumber \phi(x)|_{x=\delta x}=\phi_c(0), \\ \nonumber \text{~and~}\left(\frac{\partial \phi}{\partial x}\right)_{-\delta x}^{\delta x}=-\frac{2m}{\hbar^2}(E-V_0)2\delta x\phi_c(0), \label{eqn:newbc1}\end{eqnarray}
using which the reflection ($R(E)$) and transmission coefficients ($T(E)$) can be calculated and are given by,
\begin{equation}
    R (E) =\abs{\frac{B}{A}}^2=\frac{(m/(\hbar^2k_1)(E-V_0)2\delta x)^2}{1^2+(\frac{m}{\hbar^2k_1}(E-V_0)2\delta x)^2},
    \label{eqn:ref}
\end{equation}
\noindent and
\begin{equation}
    T (E) =\abs{\frac{C}{A}}^2=\frac{1}{1^2+(\frac{m}{\hbar^2k_1}(E-V_0)2\delta x)^2}.
    \label{eqn:tra}
\end{equation}
and the system obeys the relation, $R+T=1$. We see that the results of  ultra-short barrier matches with that of rectangular potential under limits. When the wavefunction $\phi_c(x)$ is chosen to be that of rectangular potential,
\begin{equation}
\phi_c(x)=Fe^{ik_2x}+Ge^{-ik_2x}.
\end{equation}
yields the same reflection and transmission coefficients given by Eqs \eqref{eqn:ref} and \eqref{eqn:tra}.  
This illustrate that ultra-short can be a representative of rectangular potential in its ultra-short width, the  observation which is not limited to any particular potential form. The wavenumber $k_2$ of the above equation is given by, $k_2^2=\frac{2m(E-V_0)}{\hbar^2}$.

\subsection{Transmission coefficient for ultra-short potential well}
\noindent The scattering of the quantum objects by an attractive barrier is known as Ramsauer effect \cite{liboff2003introductory}. The calculations are similar to given in the previous section, other than replacing every $V_0$ of previous calculation by -$V_0$. The coefficients can easily be derived as written below
\begin{equation}
R(E) = \abs{\frac{B}{A}}^2=\frac{(m/(\hbar^2k_1)(E+V_0)2\delta x)^2}{1^2+(\frac{m}{\hbar^2k_1}(E+V_0)2\delta x)^2},
\label{eqn:rref}
\end{equation}
\begin{equation}
T(E) = \abs{\frac{C}{A}}^2=\frac{1}{1^2+(\frac{m}{\hbar^2k_1}(E+V_0)2\delta x)^2}.
\label{eqn:rtra}
\end{equation}
\noindent with a new definition of the wave number as, $k'_2=\sqrt{\frac{2m(V_0+E)}{\hbar^2}}$. 
\begin{widetext}
\section{Results and discussions}
\subsection{Quantum properties of the ultra-short potential function}
\noindent This section presents the quantum properties of the ultra-short potential function. The results show that the proposed model is different from the rectangle potential and Dirac delta function potential. 
\textbf{Bound eigenstates:} \noindent The eigenstates of an ultra-short potential well is plotted (Fig. \ref{fig:bound}) using Eqs. \eqref{eqn:phi}, \eqref{eqn:Epm}\& \eqref{eqn:normconst}. The comparison is done by taking the area of the ultra-short potential well to be the strength of the Dirac delta function ($k_0$). From the definition of an ultra-short
\begin{equation}
\int_{-\infty}^{\infty}V_{0} f(x)\mathrm{d} x = V_{0} 2 \delta x 
\end{equation}
\noindent and 
\begin{equation}
\int_{-\infty}^{\infty}\mathrm{d}x V_0 f(x)\psi(x) = V_0\psi(0)2\delta x.
\end{equation}
\begin{figure}[h!]
    \centering
    \includegraphics[scale=1]{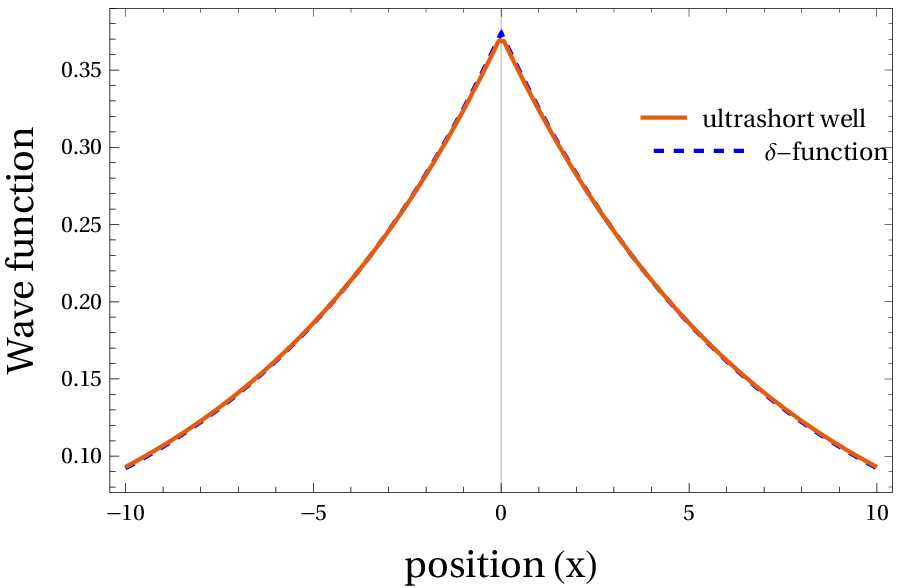}\includegraphics[scale=1]{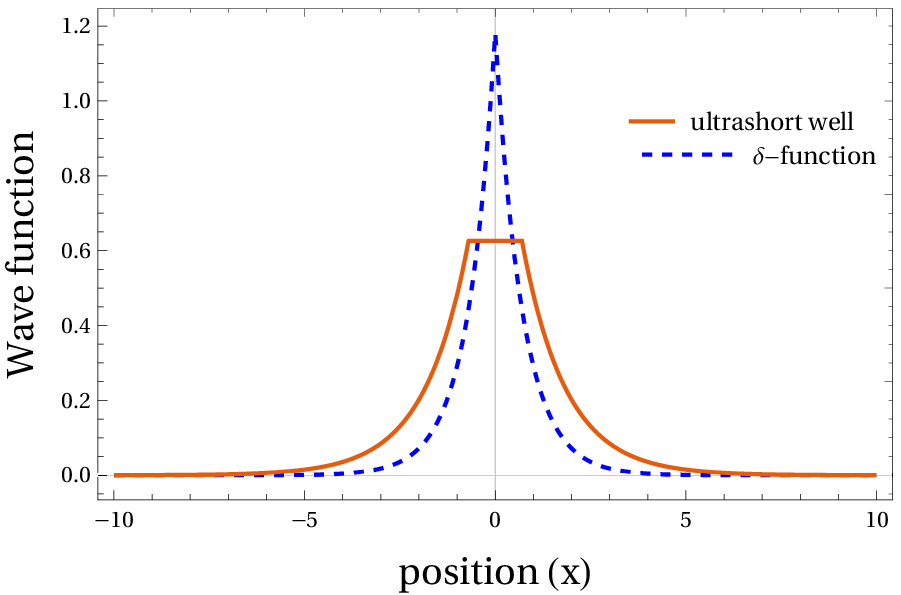}
  \caption{\centering Comparison between the eigenstates of Dirac delta and ultra-short well. The values are, $m=1$, $\hbar$=1, $V_0$=1, in (a) $\delta x$=0.07, $k_0=0.14$, in (b) $\delta x$=0.7, $k_0=1.4$.}
    \label{fig:bound}
\end{figure}
It is seen that when the width of the well is sufficiently small, eigenfunction of ultra-short potential matches that of the Dirac delta potential (Fig. \ref{fig:bound}). Being that the Dirac delta function has an infinite value which is unphysical, the above result signifies that ultra-short can serve as an alternative model to Dirac's delta function. The effect of width of the potential on bound state energy level is shown in Fig. \ref{fig:levels}. As known earlier, the results show that narrower wells have lesser binding energy. The proposed model has a single bound state like a Dirac delta function, but the energy value and wavefunction are different for both the case.
\begin{figure}[h!]
    \centering
    \includegraphics[scale=1]{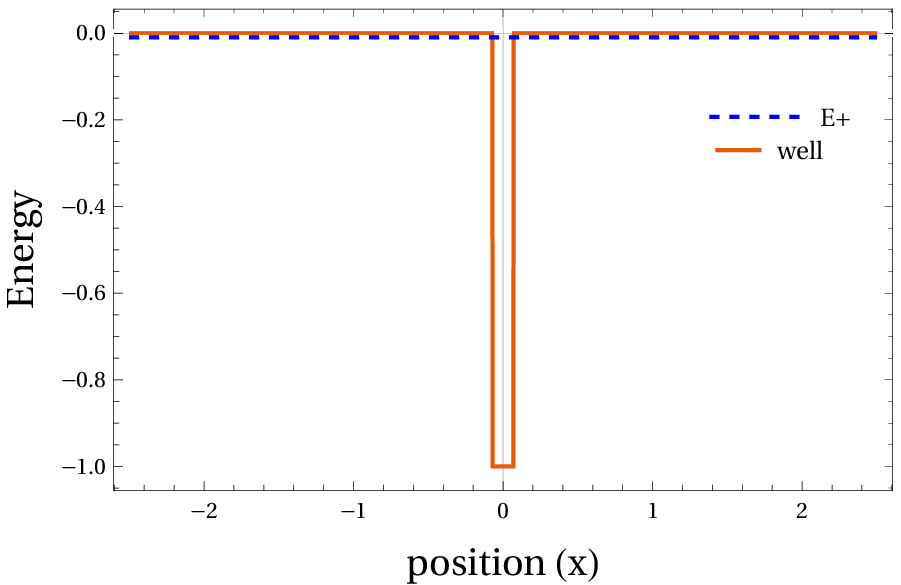}\includegraphics[scale=1]{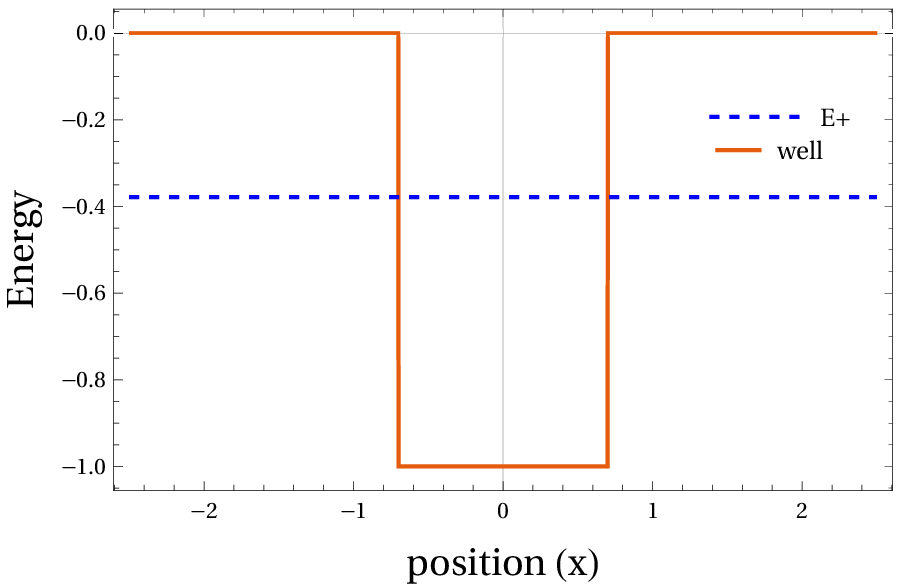}
    \centering\caption{Illustration of energy levels for values $m=1$, $\hbar$=1, $V_0$=1 for various width values: (a) $\delta x$=0.07, (b) $\delta x$=0.7. (see Fig. \ref{fig:bound} for the corresponding wave functions)}
    \label{fig:levels}
\end{figure}
\begin{figure}
    \centering
    \includegraphics[scale=1]{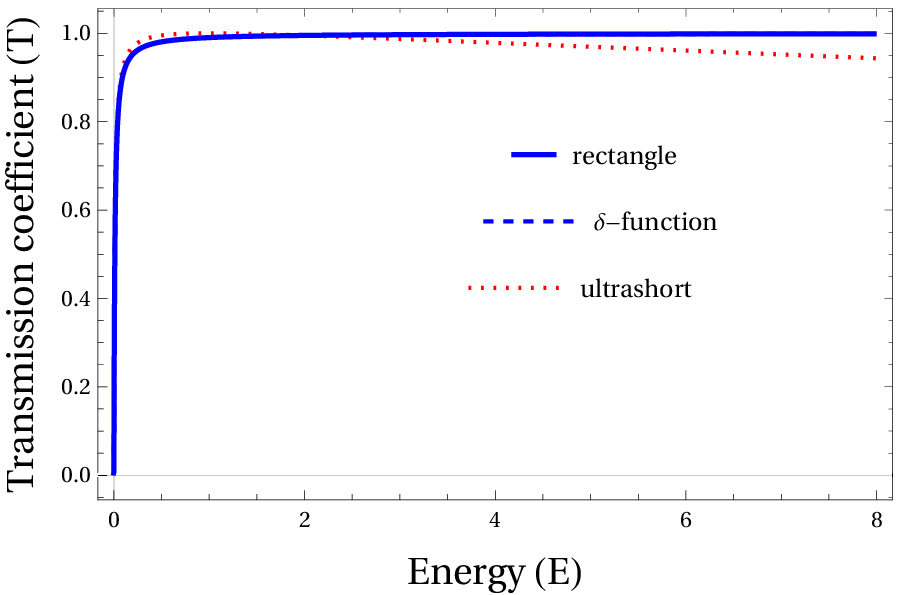}\includegraphics[scale=0.7]{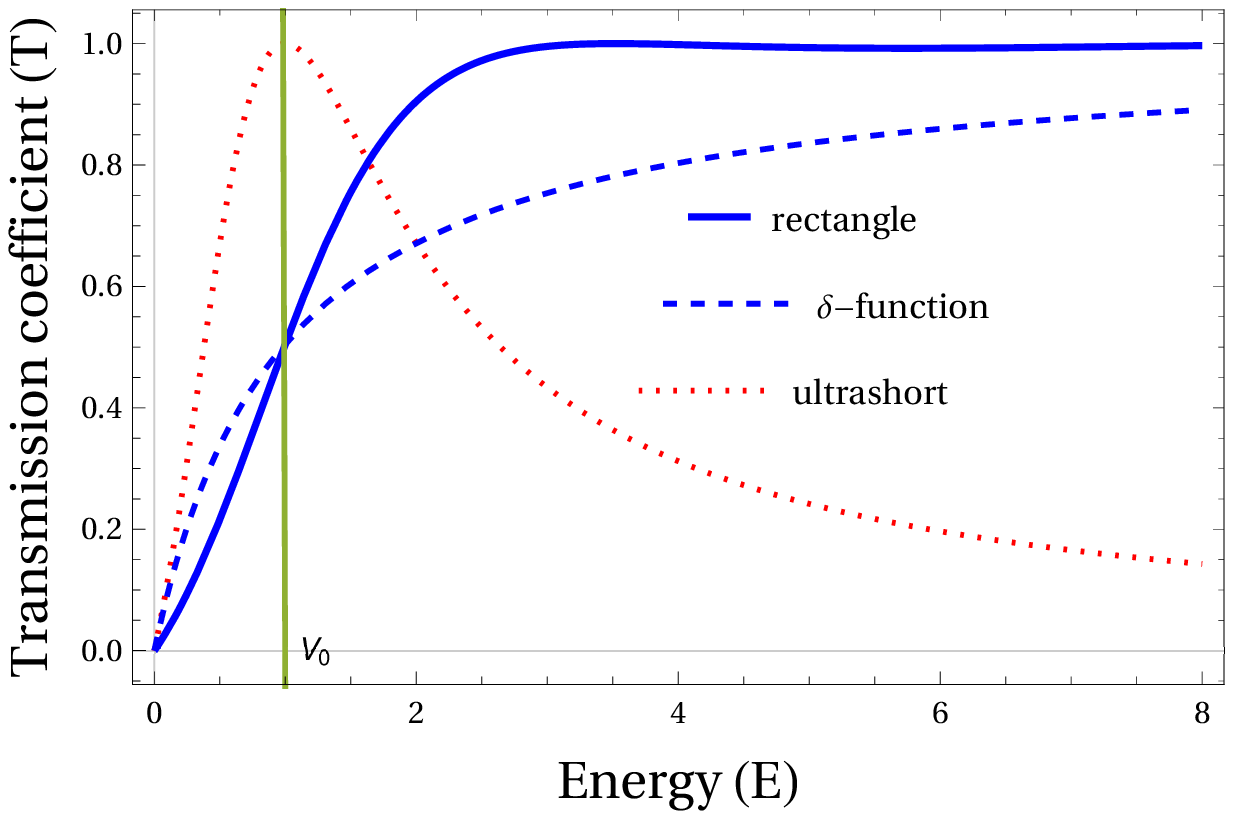}
    \caption{The plot of the transmission probability with respect to energy for ultra-short, rectangular and Dirac delta potentials. The values are $m=1$, $\hbar=1$, $V_0$=1 in a) $\delta x$=0.07, $k_0=2V_0\delta x=0.14$, in b) $\delta x$=0.7, $k_0=1.4$. The (green) vertical line gives the resonant energy of the ultra-short barrier.}
    \label{fig:tvse}
\end{figure}\\
\textbf{Unbound states: Transmission properties of the ultra-short well:} \noindent Unlike the rectangular barrier, the transmission behavior of our ultra-short potential shows resonance at $E=V_0$. 
The transmission property (Fig. \ref{fig:tvse}) can be described as follows:
\begin{itemize}
    \item The transmission probability is zero for zero energy. It increases when the energy of incoming flux increases.
    
    \item For higher energies, transmission probability decreases when the energy of incoming flux increases.
    
    \item It can be shown that the transmission behavior of the rectangular barrier converges to that of the ultra-short for smaller width values (Fig. \ref{fig:tvse}). The transmission probability of the rectangular barrier is given by \cite{liboff2003introductory},
    \begin{equation}
        T(E)=\left[\cos^2{2k_2a}+\left(\frac{k_1^2+k_2^2}{k_1k_2}\right)^2\frac{\sin^2{2k_2a}}{4}\right]^{-1}.
        \end{equation}
\end{itemize} 
\noindent When the width value `a' is small, $a=\delta x$, $\sin{2k_2a}\approx 2k_2a$ and $\cos{2k_2a}\approx1-2k_2^2a^2$. This reduction yields to,
\begin{equation}
    T=\left[1+k_1^2\delta x^2+\frac{k_2^4}{k_1^2}\delta x^2\right]^{-1}.
\end{equation}
In finite energy regime, the term $k_1^2\delta x^2$ has a negligible contribution in the transmission. After simplification we get the following
\begin{equation}
    T=\left[1+\frac{2m(E-V_0)^2}{\hbar^2E}\delta x^2\right]^{-1},
    \label{eqn:converge}
\end{equation}
\noindent which is exactly the result of an ultra-short potential (see Eq. \eqref{eqn:tra}).The representation of an ultra-short potential for an rectangular potential with low energy and width is illustrated, but the observation is not limited to any particular potential function. Fig. \ref{fig:tvse} signifies that a device which is prepared with such an ultra-short potential can have a sharp conductance with a finite energy-bandwidth around the barrier height $V_0$.\\
\textbf{Scattering probability of an ultra-short well :} \noindent The Ramsauer-Townsend scattering probability of the ultra-short, rectangular and Dirac delta potential are plotted (See Fig. \ref{fig:ramsa}). In a sufficiently small values of the width of the rectangular potential well one can reproduce the same result for ultra-short potential well.
\begin{itemize}
    \item The probability of Ramsauer scattering is zero at $E=0$. It grows upto a maxima, saturates and starts falling down. It shows that there can be no resonant transmission in ultra-short well.
    \item The saturation point is found by maximizing the functional $T(E)=\abs{\frac{C}{A}}^2=\frac{1}{1^2+(\frac{m}{\hbar^2k_1}(E+V_0)2\delta x)^2}$
 with the following constraints: 
    \begin{subequations}
     \begin{align}
        E\geq0,\\
           1\geq T(E)\geq 0,
    \end{align}
    \end{subequations}
    which yields $E=V_0$. This result is verifiable from the plot as well. The peak value of transmission is given by, $$T_{peak}=T(E=V_0)=\frac{\hbar^2}{\hbar^2+8\delta x^2mV_0}$$ as can be seen from the plot.
    \item The curve of Ramsauer scattering probability starts decreasing with increasing energy as there are no energy states to resonate the energy of the incoming object.  The analytical convergence of the rectangular well to the result of ultra-short can be shown similarly as in the earlier section.
   \end{itemize}
   \begin{figure}
    \centering
    \includegraphics[scale=1]{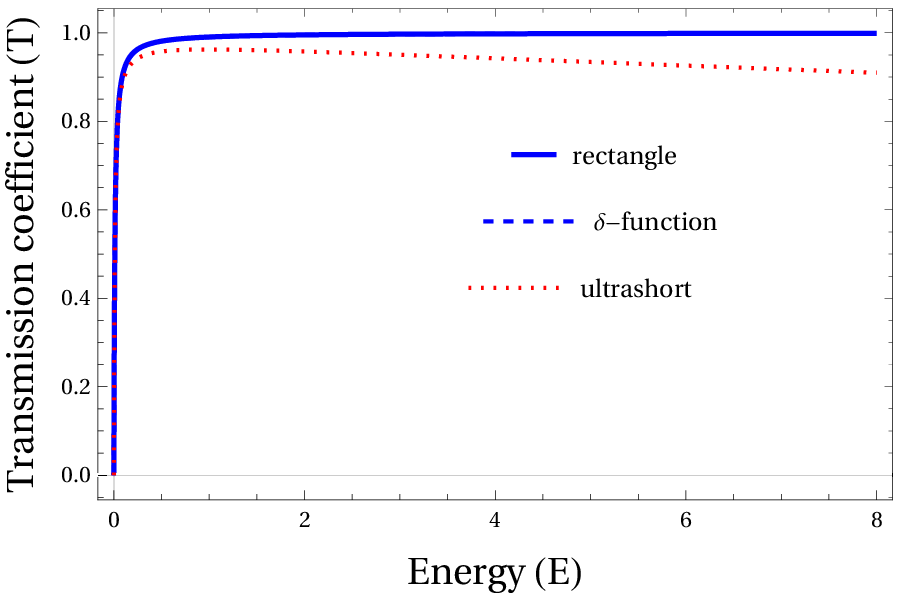}\includegraphics[scale=0.75]{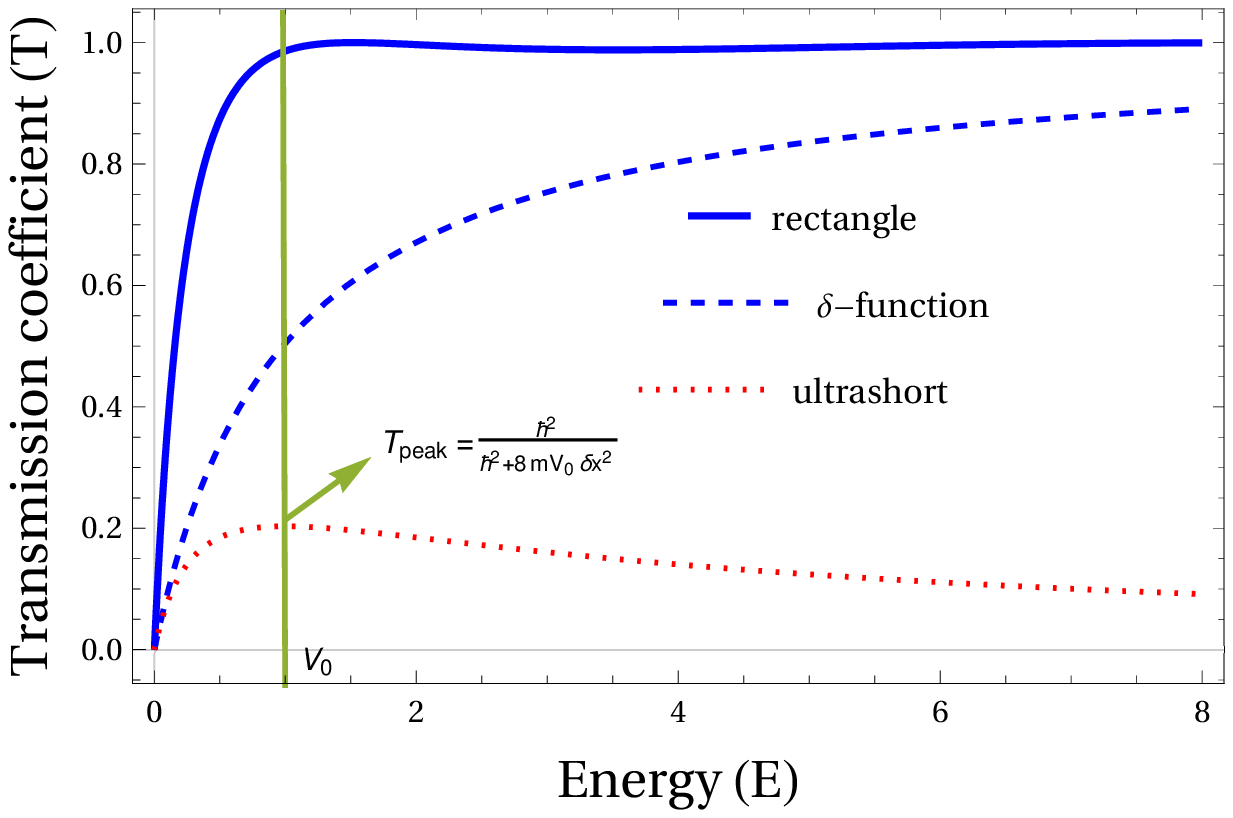}
    \caption{The plot of transmission coefficient versus energy for three potentials: ultra-short, rectangular and Dirac delta. The values are $m=1$, $\hbar=1$, $V_0=1$, a) $\delta x$=0.07, $k_0=0.14$, b) $\delta x$=0.7, $k_0$=1.4. The green vertical line signifies the energy for which there is a peak in the transmission.}
    \label{fig:ramsa}
\end{figure}
\section{The solvability of the model in the time-dependent framework}
\noindent The proposed analytical model is solvable in time-dependent framework as well. Since now, there has been no scattering models that are exactly solvable in the time-dependent framework unless when the potential involves a Dirac delta function ($V(x)=v_0\delta(x)$). Consider the time-dependent Schr\"odinger equation,
\begin{equation}
    i\hbar\frac{\partial \psi(x,t)}{\partial t} = - \frac{\hbar^2}{2m}\frac{\partial^2 \psi(x,t)}{\partial x^2}+v_0 f(x) \psi(x,t).
\end{equation}
\noindent Where $f(x)$ is 1 for a sufficiently small width around zero. We replace $v_0 f(x) \psi(x,t)$ by $v_0 f(x) \psi(0,t)$ and taking the Laplace transform, the above equation becomes
\begin{equation}
i\hbar \left[s{\tilde\psi}(x,s)-\psi(x,0)\right] =-\frac{-\hbar^2}{2m}\frac{\partial^2 {\tilde \psi}(x,s)}{\partial x^2}+v_0  f(x){\tilde \psi}(0,s).
\label{eqn:lap}
\end{equation}
\noindent The solution of the above equation in terms of Green's function is given below
\begin{equation}
{\tilde \psi}(x,s)=i\hbar \int^\infty_{-\infty} G_0(x,s|x_0) \psi(x_0,0)dx_0 + a(s) G_0(x,s|0).
\end{equation}
$G_0(x,s|x_0)$ is the Green's function of the free particle Hamiltonian. Using the boundary condition given by Eq. \ref{eqn:newbc}, the coefficient a(s) is found to be,
\begin{equation}
    a(s)=\frac{-m}{i\hbar^2}(i\hbar s-v_0)2\delta x\tilde{\psi}(0,s)\sqrt{\frac{\hbar}{2mis}}e^{-i\sqrt{\frac{2mis}{\hbar}}\abs{\delta x}}.
\end{equation}
\noindent The equation for $\tilde{\psi}(x,s)$ becomes,
\begin{equation}
{\tilde \psi}(x,s)=i\hbar \int^\infty_{-\infty}G_0(x,s|x_0)\psi(x_0,0)dx_0 - \frac{m}{i\hbar^2}(i\hbar s-v_0)2\delta x\tilde{\psi}(0,s)\sqrt{\frac{\hbar}{2mis}}e^{i\sqrt{\frac{2mis}{\hbar}}(\abs{x}-\abs{\delta x})}.
\label{eqn:psixs}
\end{equation}
after some algebra we get,
\begin{equation}
\tilde{G}_1(x,s|x_0)=G_0(x,s|x_0)- \frac{\frac{m}{i\hbar^2}(i\hbar s-v_0)2\delta x e^{i\sqrt{\frac{2mis}{\hbar}}(\abs{x}-\abs{\delta x})}\sqrt{\frac{\hbar}{2mis}}}{1+ \frac{m}{i\hbar^2}(i\hbar s-v_0)2\delta x\sqrt{\frac{\hbar}{2mis}}e^{-i\sqrt{\frac{2mis}{\hbar}}(\abs{\delta x})}}G_0(0,s|x_0),
\end{equation}
\noindent and the wavefunction can be calculated from the following expression 
\begin{equation}
{\tilde \psi}(x,s)=i\hbar \int^\infty_{-\infty}dx_0 \tilde{G}_1(x,s|x_0)\psi(x_0,0),
\end{equation}
When the asymptotic potentials are free i.e., $V(\pm\infty)=0$ the above equation simplifies to,
\begin{equation}
\tilde{G}_1(x,s|x_0)=\left(-\frac{m}{i\hbar^2}\right)\sqrt{\frac{\hbar}{2mis}}e^{i\sqrt{\frac{2mis}{\hbar}}\abs{x-x_0}}+ \frac{(i\hbar s-v_0)2\delta x\left(\frac{-m}{i\hbar^2}\sqrt{\frac{\hbar}{2mis}}\right)^2 e^{i\sqrt{\frac{2mis}{\hbar}}(\abs{x}+\abs{x_0}-\abs{\delta x})}}{1+ \frac{m}{i\hbar^2}(i\hbar s-v_0)2\delta x\sqrt{\frac{\hbar}{2mis}}e^{-i\sqrt{\frac{2mis}{\hbar}}(\abs{\delta x})}}.
\end{equation}
The result is different from that of \cite{mudra2019exact} because of the difference in the boundary conditions. The solution is presented when the asymptotic energies are free but can be extended for any potentials by writing the corresponding Green's function of the potential in the place of $G_0(x,s|x_0)$ \cite{mudra2019exact}. This propagator can be used on any initial wave packet to give the solution in Laplace domain.
\section{Transfer matrix treatment for an arbitrary potential}
\noindent An aribitrary potential can be constructed for solving a potential that consists of collection of ultrashort potential.
The wave function amplitudes are related using the boundary conditions (Eq. \ref{eqn:newbc})and the forms of the transfer matrices are derived for the different cases when $E>V(\pm\infty)$, $E<V(\pm\infty)$ and $E=V(\pm\infty)$ in the following.
\begin{figure}[h!]
    \centering
    \includegraphics[scale=1.7]{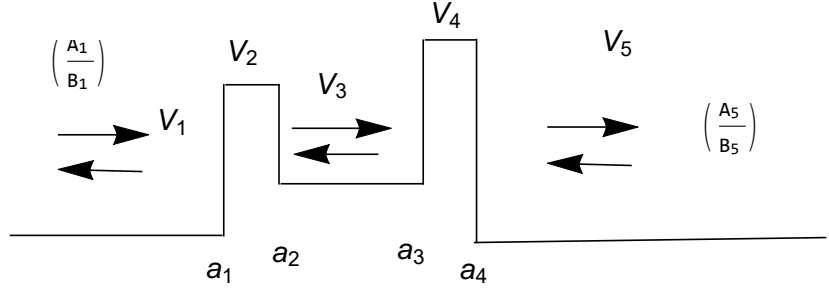}
    \caption{A schematic representation of ultra-short potentials with 5 potential regions}
    \label{fig:collscheme}
\end{figure}
Consider a potential as shown in Fig. \ref{fig:collscheme}, having 5 potential regions and 4 breakpoints. The wave function amplitudes in the different regions are related by the boundary conditions given by the Eqs. \ref{eqn:newbc1}. The relation between the coefficients can be written in the following matrix form for the scattering case as,
\begin{equation}
\left[ \begin{array}{c}
A_1 \\
B_1 \\
\end{array}\right]=
\left[A_1(a_1, V_1)\right]^
{-1}
[E(a_1,a_2,V_2,V_3)]\left[A_3(a_3, V_3)\right]^
{-1}
[E(a_2,a_3,V_4,V_5)]
\left[ \begin{array}{c}
A_5 \\
B_5\\
\end{array}\right]
\end{equation}
The matrices are given by,
\begin{equation}
A_1(V_1, a_1)=  \left[ \begin{array}{cc}
e^{ik_1a_1} &e^{-ik_1a_1} \\
-ik_1e^{ik_1a_1}& ik_1e^{-ik_1a_1}\\
\end{array}\right] ,
[E(a_1,a_2,V_2,V_3)]=  \left[ \begin{array}{cc}
e^{ik_3a_2} &e^{-ik_3a_2} \\
e^{ik_3a_2}(-ik_3+& e^{-ik_3a_2}(ik_3\\
-\frac{2m}{\hbar^2}(E-V_2)(a_2-a_1)) &-\frac{2m}{\hbar^2}(E-V_2)(a_2-a_1))
\end{array}\right],
\end{equation}
and
\begin{equation}
 [A_3(V_3, a_3)]=  \left[ \begin{array}{cc}
e^{ik_3a_3} &e^{-ik_3a_3} \\
-ik_3e^{ik_3a_3}& ik_3e^{-ik_3a_3}\\
\end{array}\right],
[E(a_3,a_4,V_4,V_5)]= \left[ \begin{array}{cc}
e^{ik_5a_4} &e^{-ik_5a_4} \\
e^{ik_5a_4}(-ik_5+& e^{-ik_5a_4}(ik_5\\
-\frac{2m}{\hbar^2}(E-V_4)(a_4-a_3)) &-\frac{2m}{\hbar^2}(E-V_4)(a_4-a_3))
\end{array}\right].
\end{equation}
The above relation can be extrapolated for multiple ultra-short potentials having N-regions and (N-1)-break points. The matrix multiplication is,
\begin{equation}
    \left[ \begin{array}{c}
A_1=1 \\
B_1 \\
\end{array}\right]=\prod_{j=0}^{\frac{N-3}{2}}
\left[A_{2j+1}(a_{2j+1}, V_{2j+1})\right]^
{-1}
[E(a_{2j+1},a_{2j+2},V_{2j+2},V_{2j+3})]
\left[ \begin{array}{c}
A_N \\
B_N=0\\
\end{array}\right],
\end{equation}
where the matrices are given by,
\begin{equation}
A_j(V_j, a_j)=  \left[ \begin{array}{cc}
e^{ik_ja_j} &e^{-ik_ja_j} \\
-ik_je^{ik_ja_j}& ik_je^{-ik_ja_j}\\
\end{array}\right] ,
[E(a_j,a_{j+1},V_{j+1},V_{j+2})]= \left[ \begin{array}{cc}
e^{ik_{j+2}a_{j+1}} &e^{-ik_{j+2}a_{j+1}} \\
e^{ik_{j+2}a_{j+1}}(-ik_{j+2}+& e^{-ik_{j+2}a_{j+1}}(ik_{j+2}\\
-\frac{2m}{\hbar^2}(E-V_{j+2})(a_{j+1}-a_j)) &-\frac{2m}{\hbar^2}(E-V_{j+2})(a_{j+1}-a_j))
\end{array}\right].
\label{transmat}
\end{equation}
The total transfer matrix is obtained by the product,
\begin{equation}
    \prod_{j=0}^{\frac{N-3}{2}}
\left[A_{2j+1}(a_{2j+1}, V_{2j+1})\right]^
{-1}
[E(a_{2j+1},a_{2j+2},V_{2j+2},V_{2j+3})]= \left[ \begin{array}{cc}
t_{11}&t_{12} \\
t_{21}&t_{22} \\
\end{array}\right].
\end{equation}
from which the transmission probability as a function of parameters can be calculated using the following simple expression \cite{gilmore2004elementary},
\begin{equation}
    T(E)=\frac{k_N}{k_1}\frac{1}{\abs{t_{11}}^2}. 
    \label{eqn:trans}
\end{equation}
Similarly for the bound state case, the coefficients can be related as given by the following equation,
\begin{equation}
    \left[ \begin{array}{c}
A_1 \\
B_1=0\\
\end{array}\right]=\prod_{j=0}^{\frac{N-3}{2}}
\left[A_{2j+1}(a_{2j+1}, V_{2j+1})\right]^
{-1}
[E(a_{2j+1},a_{2j+2},V_{2j+2},V_{2j+3})]
\left[ \begin{array}{c}
A_N \\
B_N\\
\end{array}\right]= \left[ \begin{array}{cc}
t_{11}&t_{12} \\
t_{21}&t_{22} \\
\end{array}\right]\left[ \begin{array}{c}
A_N=0 \\
B_N\\
\end{array}\right],
\end{equation}
and the bound state energies for the series can be obtained by finding the zeros of $t_{11}(E)$ \cite{gilmore2004elementary}.\\
\begin{table}[]
    \centering
  \begin{tabular}{|p{1.5cm}||p{1.5cm}|p{1.5cm}|p{4.1cm}|p{9cm}| }
 \hline
 \multicolumn{5}{|c|}{Transfer matrix for ultra-short model} \\
 \hline
\small{Cases}&\small{($A_1,B_1$)}   &\small{($A_N,B_N$)}&\small{$A_j(V_j, a_j)$} &\small{$E(a_j,a_{j+1},V_{j+1},V_{j+2})$}|\\
 \hline
 \small{$E>V$}&\small{$(1, B_1)$} &\small{$(A_N, 0)$} &\small{$$ \left[ \begin{array}{cc}
e^{ik_ja_j} &e^{-ik_ja_j} \\
-ik_je^{ik_ja_j}& ik_je^{-ik_ja_j}\\
\end{array}\right] $$}&   \small{$$ \left[ \begin{array}{cc}
e^{ik_{j+2}a_{j+1}} &e^{-ik_{j+2}a_{j+1}} \\
e^{ik_{j+2}a_{j+1}}(-ik_{j+2}+& e^{-ik_{j+2}a_{j+1}}(ik_{j+2}\\
-\frac{2m}{\hbar^2}(E-V_{j+2})(a_{j+1}-a_j)) &-\frac{2m}{\hbar^2}(E-V_{j+2})(a_{j+1}-a_j))
\end{array}\right]$$}\\
\small{$E<V$}&\small{$(A_1, 0)$} &\small{$(0, B_N)$} &\small{$$\left[ \begin{array}{cc}
e^{\kappa_ja_j} &e^{-\kappa_ja_j} \\
-\kappa_je^{\kappa_ja_j}& \kappa_je^{-\kappa_ja_j}\\
\end{array}\right]$$}&  \small{$$ \left[ \begin{array}{cc}
e^{\kappa_{j+2}a_{j+1}} &e^{-\kappa_{j+2}a_{j+1}} \\
e^{\kappa_{j+2}a_{j+1}}(-\kappa_{j+2}+& e^{-\kappa_{j+2}a_{j+1}}(\kappa_{j+2}\\
-\frac{2m}{\hbar^2}(E+V_{j+2})(a_{j+1}-a_j)) &-\frac{2m}{\hbar^2}(E+V_{j+2})(a_{j+1}-a_j))
\end{array}\right]$$}\\
\small{$E=V$}&\small{$(A_1, B_1)$} & \small{ $(A_N, B_N)$}  &\small{$$ \left[ \begin{array}{cc}
1 &a_j \\
0&1\\
\end{array}\right]$$}& \small{$$ \left[ \begin{array}{cc}
1 &a_{j+1} \\
0&1\\
\end{array}\right]$$}\\
\hline
\end{tabular}
  \caption{Transfer Matrix for the three different cases : $E>V$, $E<V$, $E=V$.}
    \label{tab:transmat}
\end{table}

\noindent\textbf{Calculating quantum properties using the transfer matrix approach }\\
\noindent The transmission coefficient is calculated using the proposed scheme for the rectangular potential and is compared with the analytical formula. The Fig. \ref{fig:scattering} shows a slight variation. For higher accuracy, more number of ultrashort potentials can be employed in the collection. The matrix multiplication and evaluation are done using MATHEMATICA\cite{Mathematica}. For the case of wells, the eigenenergies are presented in \autoref{tab:eigenvalues} using the scheme and compared with that from the analytical formula. The energy values show slight variation from the actual number. It is shown that the variation can be minimized by using more number of potentials for the same width (Shown in row 4 and 5 of \autoref{tab:eigenvalues}). The well is characterized by the parameters : the potential $V=V_{asymptote}-V_{inside}$ and $a$ is the width of the well. The eigenfunction match qualitatively with that of the available result (Fig. \ref{fig:wavefunction}). The advantage of the presented scheme is that for N-energy regions, one has to multiply a total of N-1 matrices. On the other hand, the piece-wise constant construction requires a matrix multiplication of N+4 matrices at the best \cite{gilmore2004elementary}. The piece-wise linear construct involves matrix multiplication involves 2(N-1) matrix multiplication \cite{lui1986exact}. The transmission coefficient of the rectangular potential is calculated using the code and compared with the analytical formula (Fig. \ref{fig:scattering}). And for more accurate results, large number of collections have to be used. The table \ref{tab:comparison} gives the comparison between the other schemes. Computational complexity gives the number of matrix multiplications for N-regions. The solubility of the model in time-dependent framework is also considered in the comparison. Some of the schemes allow detailed analysis of the quantum properties by adding 1 model potential at a time.

\begin{figure}[h!]
    \centering
    \includegraphics[scale=1]{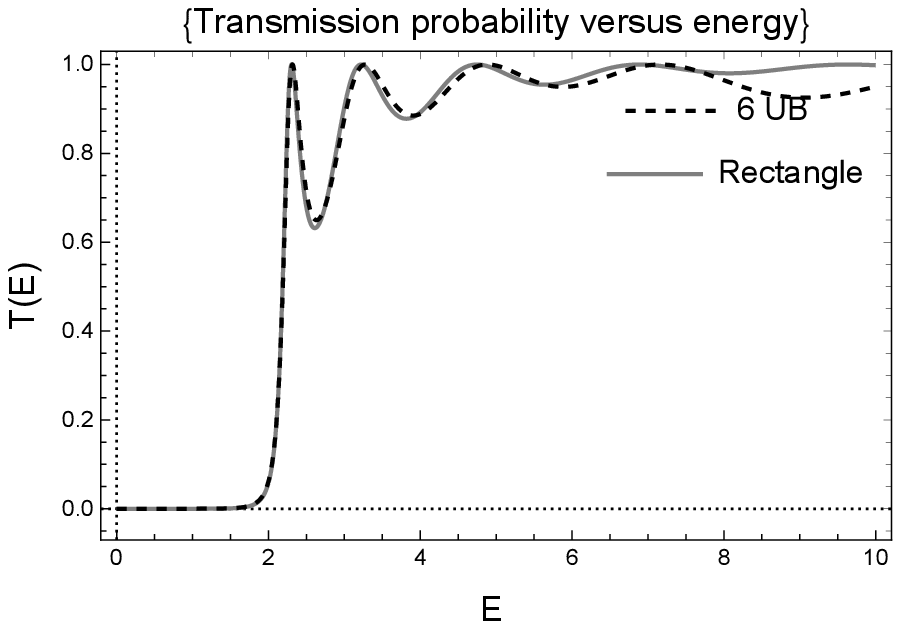}\includegraphics[scale=1]{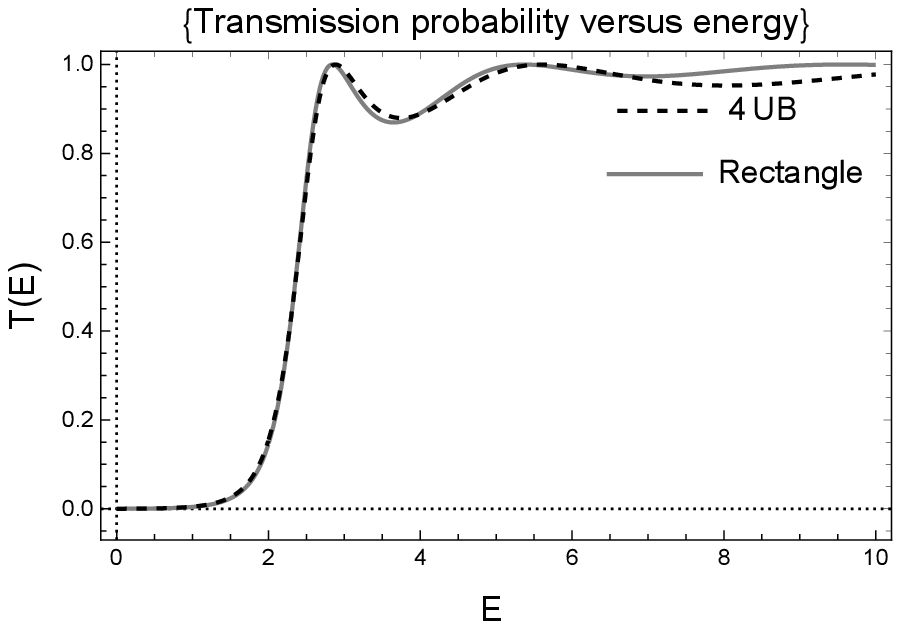}
    \caption{The plot of transmission probability versus energy for a rectangle produced using 6 and 4 ultra-short potentials. The parameters are : $m=1, \hbar=1, \delta x=0.02, V_1=V_N=0, V_2=V_3..=2, a_3-a_1=a_5-a_3..=0.8 $, which is equivalent to a rectangle of $V_0=2$ and a) $2a=2.42$ b) $2a=4.02$. It is seen that when the number of employed ultra-short potential increases, the accuracy extends to more higher energies.}
    \label{fig:scattering}
\end{figure}
\begin{table}[]
    \centering
   
\begin{tabular}{|p{1.5cm}||p{1.5cm}|p{1.5cm}|p{6.5cm}|p{6.5cm}| }
 \hline
 \multicolumn{5}{|c|}{Energy eigenvlaues calculation} \\
 \hline
\small{Number of UW}& \small{parameters ($V,a$)}&\small{No. of bound states} &\small{$E_n$ using N-wells}   &\small{$E_n$ for rectangle}\\
 \hline
 \small{1}&\small{(20,0.05)} &\small{1} &\small{\{19.5235\}}&   \small{\{19.5161\}}\\
  \small{2}&\small{(20,1.05)} &\small{3} &\small{\{ 2.96496, 10.6123, 19.5963\}}&   \small{\{2.61562, 10.0498, 19.5865\}}\\
   \small{3}&\small{(20,2.05)} &\small{5} &\small{\{0.954625, 3.90531, 8.12712, 14.2251, 19.6332\}}&   \small{\{0.87960, 3.49579, 7.76616, 13.4719, 19.622\}}\\
\small{4}&\small{(20,3.05)} &\small{7} &\small{\{0.467598, 1.872, 4.32418, 7.16934, 11.2257, 15.8682, 19.6558\}}&   \small{\{
 0.435195, 1.73706, 3.89332, 6.87898, 10.6445, 15.0739, 19.6437\}}\\
\small{5}&\small{(20,3.05)} &\small{7} &\small{\{0.45786, 1.82878, 4.10815, 7.38259, 10.84667, \
15.57358, 19.98671\}}&   \small{\{
 0.435195, 1.73706, 3.89332, 6.87898, 10.6445, 15.0739, 19.6437\}}\\
\small{5}&\small{(20,4.05)} &\small{9} &\small{\{0.27667, 1.10666, 2.49353, 4.56035, 6.66904, 9.67554, 13.0808, 16.8106, 19.6713\}}&   \small{\{0.258774, 1.0341, 2.32281, 4.119, 6.41269, 9.18687, 12.41, 16.0097, 19.6585\}}\\
\hline
\end{tabular}
   \caption{Comparison of eigenvalues between analytical formula with the transfer matrix result}
    \label{tab:eigenvalues}
\end{table}
\begin{figure}
    \centering
    \includegraphics[scale=1]{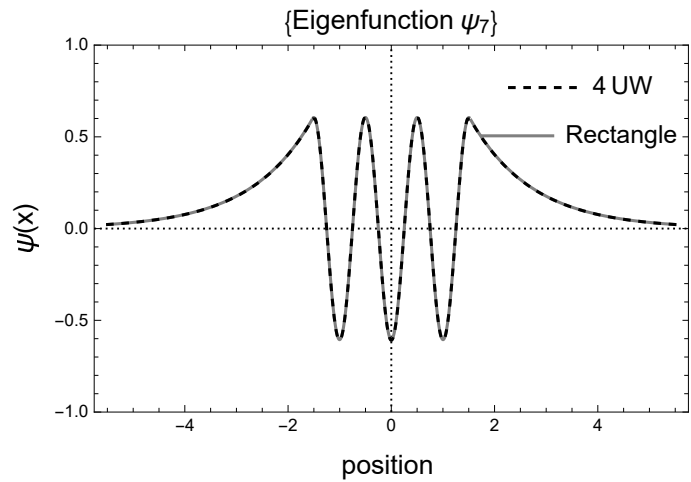}
    \caption{Comparison of the 6-th eigenfunction between collection of 4 ultra-short wells and the corresponding rectangular barrier. The parameters are $V=20, a=3.05, En=19.6558$. The energies are taken to be same to check the quantitative equality.}
    \label{fig:wavefunction}
\end{figure}
\begin{table}[h!]
    \centering
   \begin{tabular}{|p{3cm}||p{3cm}|p{3cm}|p{3cm}|p{3cm}|p{3cm}|}
 \hline
 \multicolumn{6}{|c|}{Comparison with other analytical schemes}\\
 \hline
Models&Piece-wise constant   &Piece-wise linear&rectangle &Dirac delta potentials&Our model\\
 \hline
 Computational complexity&4N+4/N+4 &2N-2 &2N-2&N-4&N-4\\
Solubility in time-domain&numerical&numerical&numerical&  exact&  Laplace domain
solution\\
Analysis friendly&0 & 0  &1& 1& 1\\
\hline
\end{tabular}
    \caption{Comparison of the ultrashort scheme with other similar schemes}
    \label{tab:comparison}
\end{table}
\section{Conclusion and Future Scope}
\noindent A new analytical scheme is introduced for solving one-dimensional Schr\"odinger equation. For building the collection scheme, a new solvable potential is introduced. The model potential has desirable properties of an analytical model. The scheme can be used to calculate the transmission probability, energy eigenvalues and the eigenfunctions for any potential. It is seen that the present method has computational advantage due to less number of matrix multiplications and also due to the simplicity in the functional form of the wavefunction. The analytical scheme can be given for time-dependent case as well using the collection\cite{saravanan2019exact}. The model can be extended to time-dependent and coupled-state systems that leads to understanding in molecular systems, chemical physics \cite{sebastian1992theory,chakraborty2009multi,gerber1974properties}. The properties of an arbitrary potential with arbitrary effective mass function can be calculated by using the Mathematica code given in the supplement.

\end{widetext}

\begin{thebibliography}{99}
\bibitem{balantekin1998quantum}
A. B. Balantekin and N. Takigawa, Rev. Mod. Phys. \textbf{70}, 77 (1998).
\bibitem{gamow1928quantentheorie}
G. Gamow, Zeits. f. Phys. \textbf{51}, 204 (1928).
\bibitem{forbes2011transmission}
R. G. Forbes and J. H. B. Deane, in \textit{Proc. Roy. Soc. Lond. A}, Vol. 467 (The Royal Society, 2011) pp. 2927–2947.
\bibitem{datta1990quantum}
S. Datta and M. J. McLennan, Rep. Prog. Phys. \textbf{53}, 1003 (1990).
\bibitem{luscombe1990models}
J. H. Luscombe and W. R. Frensley, Nanotechnology \textbf{1}, 131 (1990).
\bibitem{harris1989electronic}
J. J. Harris, J. A. Pals, and R. Woltjer, Rep. Prog. Phys. \textbf{52}, 1217 (1989).
\bibitem{kronig1931quantum}
R. L. Kronig and W. G. Penney, in \textit{Proc. Roy. Soc. Lond. A}, Vol. 130 (The Royal Society, 1931) pp. 499–513.
\bibitem{domany1983solutions}
E.  Domany, S.  Alexander, D.  Bensimon, and  L.  P. Kadanoff, Phys. Rev. B \textbf{28}, 3110 (1983).
\bibitem{chang1974resonant}
L. L. Chang, L. Esaki,  and R. Tsu, Appl. Phys. Lett. \textbf{24}, 593 (1974).
\bibitem{cahay1989electron}
M. Cahay, J. P. Kreskovsky,  and H. L. Grubin, Solid-State Electron. \textbf{32}, 1185 (1989).
\bibitem{osorio1988bound}
F. A. Osorio, M. Degani, and O. Hipólito, Phys. Rev. B \textbf{37}, 1402 (1988).
\bibitem{gerber1974properties}
R. B. Gerber and N. C. Rosenbach, Phys. Rev. A \textbf{9}, 301 (1974).
\bibitem{gilmore2004elementary}
R. Gilmore, Elementary quantum mechanics in one dimension (JHU Press, 2004).
\bibitem{ando1987calculation}
Y. Ando and T. Itoh, J. Appl. Phys. \textbf{61}, 1497 (1987).
\bibitem{lui1986exact}
W.  W.  Lui  and  M.  Fukuma, J.  Appl.  Phys. \textbf{60}, 1555 (1986).
\bibitem{chakraborty2009multi}
A. Chakraborty, Mol. Phys. \textbf{107}, 2459 (2009).
\bibitem{saravanan2019exact}
 R. Saravanan and A. Chakraborty, “Exact time-domainsolution of the Schr ̈odinger equation for a new scattering model,” Unpublished.
\bibitem{griffiths2016introduction}
D. J. Griffiths, \textit{Introduction to quantum mechanics}, 2nd ed. (Person Education, 2016).
\bibitem{liboff2003introductory}
 R.  L.  Liboff, \textit{Introductory quantum mechanics},  2nd  ed.(Addison-Wesley, 1991).
 \bibitem{mudra2019exact}
 S. Mudra and A. Chakraborty, Physica Scripta  (2019).
 \bibitem{Mathematica}
 Wolfram Research, Inc., Mathematica, Version 12.0, Champaign, IL (2019).
 \bibitem{sebastian1992theory}
K. L. Sebastian, Phys. Rev. A \textbf{46}, R1732 (1992).

\end{thebibliography}
\end{document}